\documentclass[fleqn,10pt]{SelfArx} 

\usepackage{lipsum} 

\definecolor{color1}{RGB}{0,0,90} 
\definecolor{color2}{RGB}{0,20,20} 

\usepackage{hyperref} 
\hypersetup{hidelinks,colorlinks,breaklinks=true,urlcolor=color2,citecolor=color1,linkcolor=color1,bookmarksopen=false,pdftitle={Title},pdfauthor={Author}}

\JournalInfo{J. Vis. Exp. (87), e51224 5/30/2014} 
\Archive{jove.come/video/51224\\ chistera-qscale.eu/jove.html} 

\PaperTitle{Quantum state engineering of light with continuous-wave optical parametric oscillators} 

\Authors{Olivier Morin\textsuperscript{1}, Jianli Liu\textsuperscript{1}, Kun Huang\textsuperscript{1,2}, Felippe Barbosa\textsuperscript{3}, Claude Fabre\textsuperscript{1}, Julien Laurat\textsuperscript{1}*} 
\affiliation{\textsuperscript{1}\textit{Laboratoire Kastler Brossel, Universit\'{e}
Pierre et Marie Curie, Ecole Normale Sup\'{e}rieure, CNRS, 4 place
Jussieu, 75252 Paris Cedex 05, France}} 
\affiliation{\textsuperscript{2}\textit{State Key Laboratory of Precision Spectroscopy, East China Normal University, Shanghai 200062, China}} 
\affiliation{\textsuperscript{2}\textit{Instituto de F\'{i}sica, Universidade de S\~{a}o Paulo
P. O. Box 66318, S\~{a}o Paulo, SP 05314- 970, Brazil
}} 
\affiliation{*\textbf{Corresponding author}: julien.laurat@upmc.fr} 

\Keywords{Quantum optics, Quantum state engineering, Optical parametric oscillator, Squeezed vacuum, Single photon, Coherent state superposition, Homodyne detection} 
\newcommand{\keywordname}{Keywords} 

\Abstract{
The ability to engineer the quantum state of traveling optical fields is a
central requirement for quantum information science and technology, including
quantum communication, computing and metrology. In this video article, we
describe the reliable generation of non-Gaussian states, including
single-photon states and coherent state superpositions, using a conditional
preparation method operated on the non-classical light emitted by optical
parametric oscillators. Type-I and type-II phase-matched OPOs operated below
threshold, i.e. single-mode or two-mode squeezed vacuum sources, are considered
and common procedures, such as the required frequency filtering or the
high-efficiency quantum state characterization by homodyning, are detailed. The
reported method enables a high fidelity with the targeted state and the
generation of the state in a well-controlled spatiotemporal mode, a crucial
feature for their use in subsequent protocols.\\

\textit{The video of this article can be found at jove.come/video/51224 or chistera-qscale.eu/jove.html}}

\usepackage[T1]{fontenc}
\usepackage[utf8]{inputenc}
\usepackage{authblk}

\usepackage{graphicx}
\usepackage{hyperref}
\usepackage{amsmath}
\usepackage{epstopdf}
\usepackage{bbm}

\begin{document}

\maketitle

\section{Introduction}
The ability to engineer the quantum state of traveling optical fields is a central requirement for quantum information science and technology \cite{1,2}, including quantum communication, computing and metrology. Here, we discuss the preparation and characterization of some specific quantum states using as a primary resource the light emitted by continuous-wave optical parametric oscillators \cite{3,4} operated below threshold. Specifically, two systems will be considered, a type-II phase-matched OPO and a type-I OPO , enabling respectively the reliable generation of heralded single-photons and of optical coherent state superpositions (CSS), i.e. states of the form $|\alpha\rangle-|-\alpha\rangle$. These states are important resources for the implementation of a variety of quantum information protocols, ranging from linear optical quantum computation \cite{6} to optical hybrid protocols \cite{5,7}. Significantly, the method presented here permits obtaining a low admixture of vacuum and the emission into a well-controlled spatiotemporal mode.

\begin{figure}[t!]
\centerline{\includegraphics[width=1\linewidth]{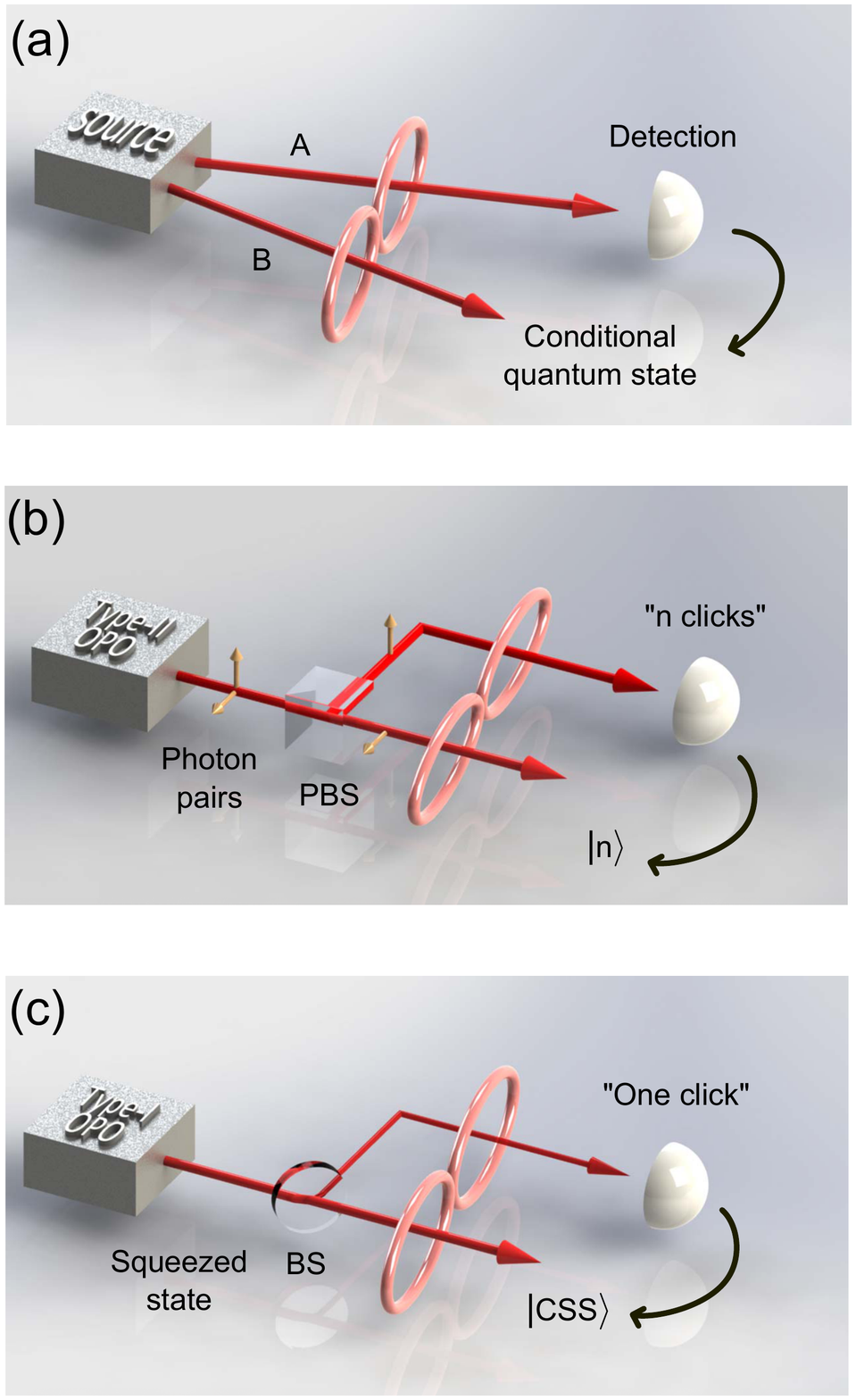}}
\caption{\textbf{Conditional state preparation technique}. \hspace{0.8cm}(a) Conceptual scheme. (b) Conditional preparation of single-photon state from orthogonally-polarized photon pairs (type-II OPO) separated on a polarizing beam splitter. (c) Conditional preparation of a coherent state superposition by subtracting a single-photon from a squeezed vacuum state (type-I OPO).}
\label{fig1}
\end{figure}

Generally speaking, quantum states can be classified as Gaussian states and non-Gaussian states according to the shape of the quasi- probability distribution in phase space called the Wigner function W(x, p) \cite{8}. For non-Gaussian states, the Wigner function can take negative values, a strong signature of non-classicality. Single-photon or coherent state superpositions are indeed non-Gaussian states.
An efficient procedure for generating such states is known as the conditional preparation technique, where an initial Gaussian resource is combined with a so-called non-Gaussian measurement such as photon counting \cite{9,10,11,12,13}. This general scheme, probabilistic but heralded, is sketched on Figure \ref{fig1}a. By measuring one mode of a bipartite entangled state, the other mode is projected into a state that will depend on this measurement and on the initial entangled resource \cite{12,13}.

What are the required resource and heralding detector needed to generate the aforementioned states? Single-photon states can be generated using twin beams, i.e. photon-number correlated beams. The detection of a single-photon on one mode then heralds the generation of a single- photon on the other mode \cite{9,10,14,15}. A frequency-degenerate type-II OPO \cite{16,17,18,19} is indeed a well-suited source for this purpose. Signal and idler photons are photon-number correlated and emitted with orthogonal polarizations. Detecting a single-photon on one polarization mode projects the other one into a single-photon state, as shown in Figure \ref{fig1}b.

Concerning coherent state superpositions, they can be generated by subtracting a single-photon from a squeezed vacuum state \cite{20} obtained either by pulsed single-pass parametric down-conversion \cite{11,21} or by a type-I OPO \cite{22,23}. The subtraction is performed by tapping a small fraction of the light on a beam-splitter and detecting a single-photon in this mode (Figure \ref{fig1}c). A squeezed vacuum is a superposition of even photon- number states, thus subtracting a single-photon leads to a superposition of odd photon-number states, which has a high fidelity with a linear superposition of two coherent states of equal and small amplitude. For this reason, the name Schr\"{o}dinger kitten has sometimes been given to this state.

The general procedure for generating these states is thus similar, but differs by the primary light source. Filtering of the heralding path and detection techniques are the same whatever the type of OPO used. The present series of protocols detail how to generate these two non- Gaussian states from continuous-wave optical parametric oscillators and how to characterize them with high efficiency.

\section{Protocol}

\subsection{Optical Parametric Oscillator}
\begin{itemize}
\item Build a 4 cm long semimonolithic linear cavity (for improved mechanical stability and reduced intracavity losses). The input mirror is directly coated on one face of the nonlinear crystal.
\item Choose an input coupler reflection of 95\% for the pump at 532 nm and high-reflection for the signal and idler at 1064 nm. Inversely, choose the output coupler to be highly reflective for the pump and of transmittance T=10\% for the infrared. The free spectral range of the OPO is equal to $\Delta\omega$ = 4.3 GHz and the bandwidth is around 60 MHz. Make the cavity triply resonant, i.e. for the pump and for the down-converted fields.
\item Use a KTP crystal for the type-II OPO system or a PPKTP crystal for the type-I OPO. Temperature-stabilize the crystals at their phase- matching temperatures.
\item Use as laser source a continuous-wave frequency doubled Nd:YAG laser. Pump the OPO at 532 nm and use the infrared light, after spatial filtering by a high-finesse cavity (mode cleaner), as a local oscillator (LO) for the homodyne detection.
\item Achieve the mode-matching between the pump and the cavity mode.
\item Lock the cavity length on the pump resonance by the Pound-Drever-Hall technique. For this purpose, apply a 12 MHz electro-optic
modulation to the pump and detect the light back-reflected from the cavity with an optical isolator.
\end{itemize}

\subsection{Conditional Preparation: filtering the Heralding Path}
\begin{itemize}
\item Separate the OPO output into two modes. One corresponds to the heralding mode, while the other one is the heralded state that will be detected by the homodyne detection.
\item Guide the heralding mode towards the single-photon detector. Specifically, for the type-II OPO, separate the orthogonal signal and idler modes by a polarized beam splitter (PBS). For the type-I OPO, tap out a small fraction (3\%) of the squeezed vacuum by a beam splitter (BS).
\item Filter the heralding mode to remove the frequency non-degenerate modes due to the OPO cavity. For an OPO, the output indeed contains many pairwise correlated but spectrally separated modes, $\omega_0+n\Delta\omega$ and $\omega_0-n\Delta\omega$ where $n$ is an integer. To generate a heralded state at the carrier frequency, it is necessary to filter out all of these non-degenerate modes.
\begin{itemize}
\item Use first an interferential filter with a bandwidth of 0.5 nm.
\item Add a homemade linear Fabry-Perot cavity with a free spectral range of 330 GHz and a bandwidth of 300 MHz (length around 0.4
mm and finesse around 1000). The cavity bandwidth is chosen to be larger than the one of the OPO and the free spectral range to be
larger than the frequency window of the interferential filter.
\item Achieve at least an overall 25 dB rejection of the non-degenerate modes.
\end{itemize}
\item Lock the filtering Fabry-Perot cavity by the dither-and-lock technique.
\begin{itemize}
\item For this purpose, inject a backward propagating auxiliary beam via an optical switch and reject it at the entrance of the filtering cavity by
an optical isolator. Detect the light at the output.
\item  Lock the cavity during 10 msec and start after the measurement period for 90 msec with the auxiliary-beam off.
\end{itemize}
\item Detect the filtered heralding mode by a single-photon detector during the measurement period. A superconducting single-photon detector (SSPD) is used to limit the amount of dark noise (few Hz), which otherwise would degrade the fidelity of the conditional state.
\end{itemize}

\subsection{Quantum State Tomography by Homodyne Detection}
\begin{itemize}
\item Detect the heralded state with a balanced homodyne detection composed of a 50/50 beam splitter where the field to characterize and a strong continuous-wave local oscillator (LO, 6 mW) are brought to interfere, and a pair of high quantum efficiency InGaAs photodiodes.
\item In order to align the detection, inject into the OPO cavity a bright auxiliary beam at 1064 nm and mode match this mode with the LO mode. Achieve a fringe visibility close to unity. Any mode mismatch quadratically translates into detection losses.
\item Check the homodyne detection properties. With a LO power of 6 mW, the shot noise limit (SNL) is flat up to 50 MHz. It is more than 20 dB above the electronic noise at low analysis frequency (MHz), 16 dB above at the analysis frequency of 50 MHz. This distance is a critical parameter as it translates into losses in the detection (a 10 dB (20 dB) distance translates into a 10\% (1\%) effective loss) \cite{24}.
\item For every detection event from the single-photon detector, record the homodyne photocurrent with an oscilloscope with a sampling rate of 5 Gs/sec during 100 nsec. Sweep the LO phase with a PZT-mounted mirror during the measurement.
\item Filter each recorded segment with a given temporal mode function to obtain at each successful preparation a single quadrature value of the conditional state. The optimal mode function for low gain is close to a double-sided exponential function \cite{25} with a decay constant equal to the inverse of the OPO bandwidth. The optimal mode can also be found by using a eigenfunction expansion of the autocorrelation function \cite{26}.
\item Accumulate measurements (50000 are required for the tomography) and post-process the data with a maximum-likelihood algorithm \cite{27}. This procedure enables reconstruction of the density matrix of the heralded state and the corresponding Wigner function \cite{8}.
\end{itemize}

\subsection{Conditional Preparation of single Photon State with a type-II OPO}
\begin{itemize}
\item Pump the type-II OPO far below threshold (1 mW here for a 80 mW threshold) to have a very low probability of multiphoton pairs.
\end{itemize}

\subsection{Conditional Preparation of Coherent State Superposition with a type-I OPO}
\begin{itemize}
\item Check the squeezed vacuum generated by the OPO close to threshold with a spectrum analyzer. The measured noise spectra are shown in Figure \ref{fig3}.
\item Operate the OPO at a pump power enabling observation of around 3 dB of squeezing at low sideband frequencies (few MHz).
\item In the homodyne measurement, the phase information is important for phase-dependent states such as the CSS state. Scan the LO phase
with a 10 Hz sawtooth wave with a duty cycle of 90\% (corresponding to the 90 msec of measurement period and 10 msec of locking period.)
Synchronize the sweep to make sure that during the measurement period, there is a single one-directional sweep of the PZT-mounted mirror.
\item Use the homodyne signal to measure the variance and then infer the phase of the measured quadrature.
\end{itemize}

\section{Representative Results}
\subsection{For the type-II OPO and the generation of high-fidelity single photon state}
\begin{figure}[ht!]
\centerline{\includegraphics[width=0.9\linewidth]{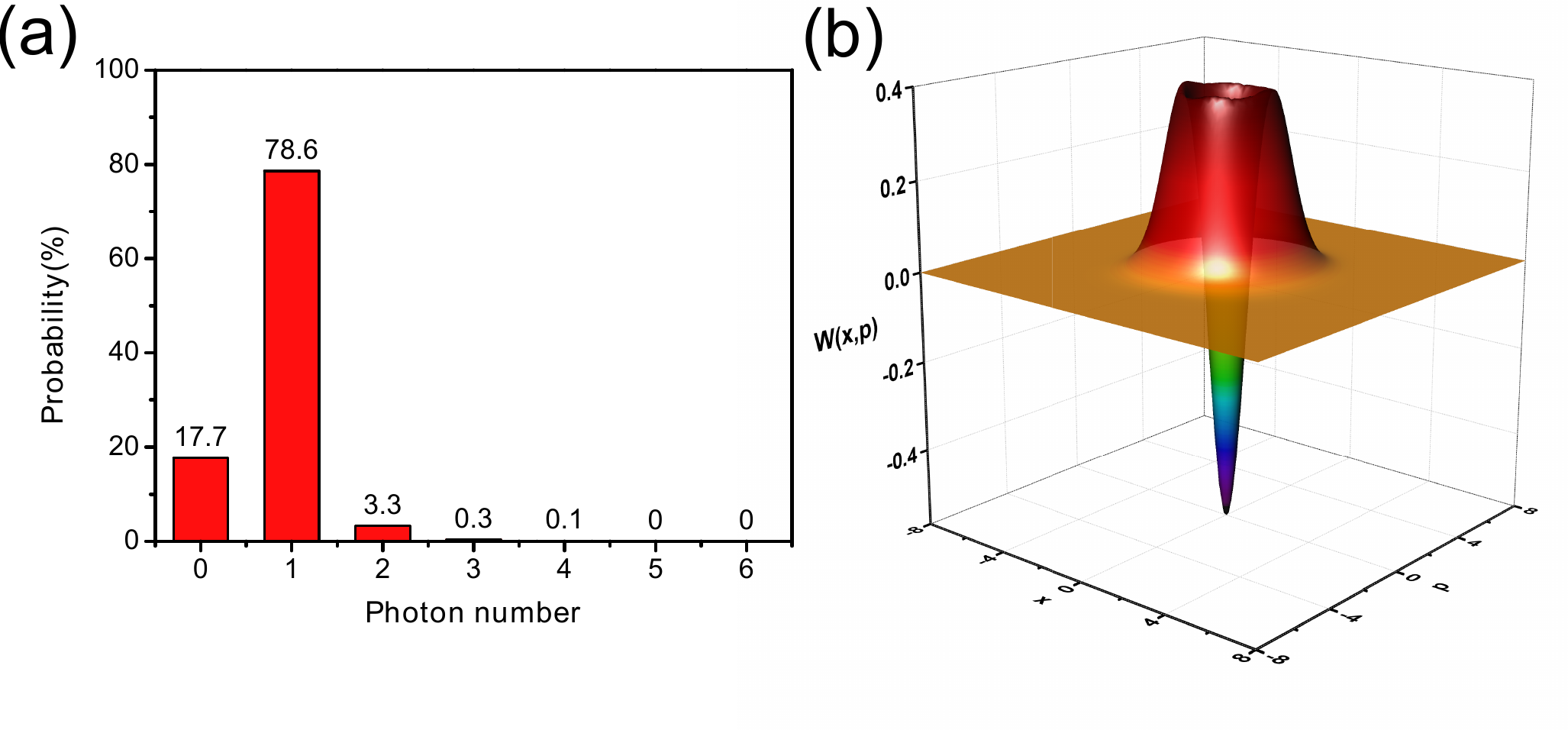}}
\caption{\textbf{High-fidelity single-photon state.} (a) Diagonal elements of the reconstructed density matrix without correction from detection losses. (b) Corresponding Wigner function. $x$ and $p$ denote quadrature components.}
\label{fig2}
\end{figure}

The tomographic reconstruction of the heralded state is shown in Figure \ref{fig2}, where the diagonal elements of the reconstructed density matrix and the corresponding Wigner function are displayed. Without any loss corrections, the heralded state exhibits a single-photon component as high as 78\%. By taking into account the overall detection losses (15\%), the state reaches a fidelity of 91\% with a single-photon state. The two-photon component, which results from multi-photon pairs creating by the down conversion process, is limited to 3\%.

\subsection{For the type-I OPO and the generation of CSS state}
\begin{figure*}[ht!]\centering
\includegraphics[width=0.9\linewidth]{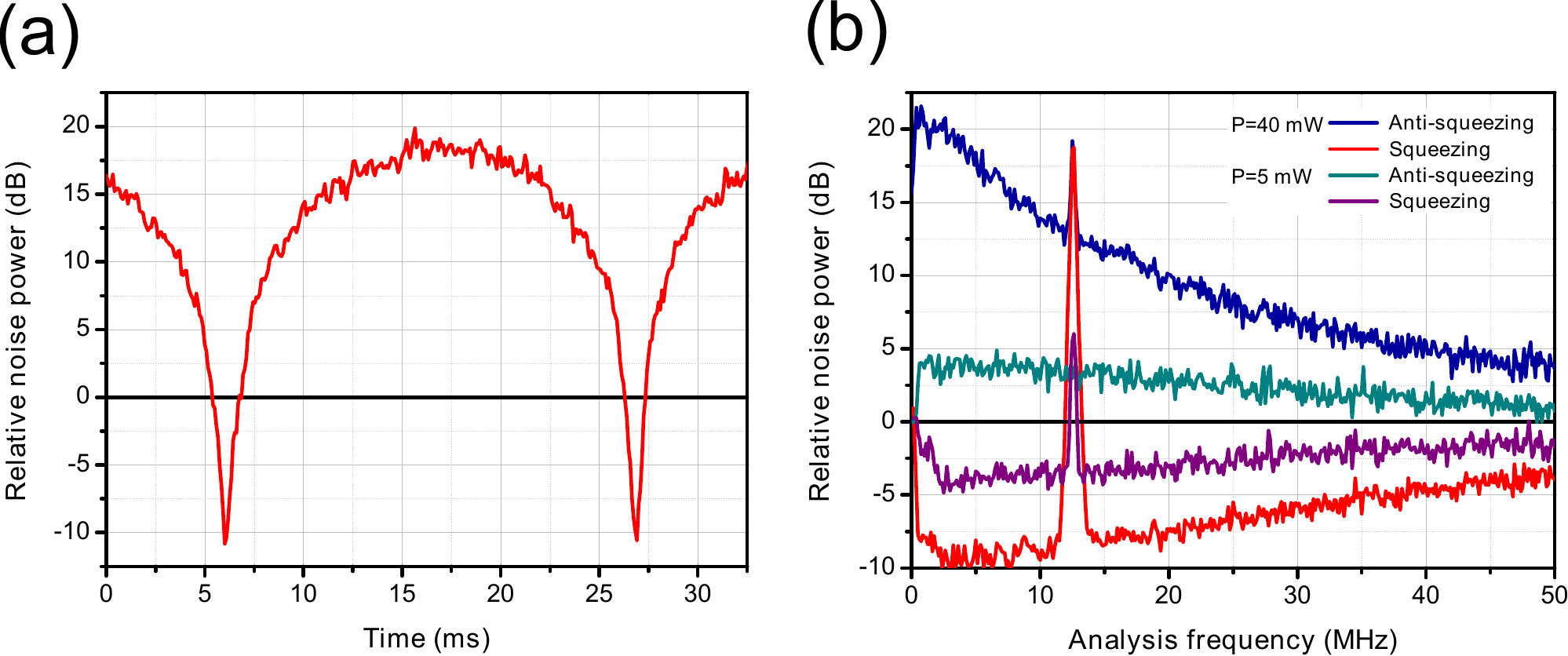}
\vspace{0.3 cm}\caption{ \textbf{Measured noise spectra of squeezed vacuum states generated by the type-I PPKTP OPO.} All the data are recorded by a spectrum analyzer with a resolution bandwidth of 300 kHz and a video bandwidth of 300 Hz. Spectra are normalized to the shot noise limit. (a) Noise variance as a function of the local oscillator phase, at a pump power of 40 mW and an analysis frequency of 5 MHz. (b) Broadband squeezing up to 50 MHz for a pump power of 5 mW and a pump power of 40 mW. The peak at 12 MHz results from the electro-optic modulation used to lock the cavities.}
\label{fig3}
\end{figure*}

\begin{figure}[t!]
\centerline{\includegraphics[width=0.95\linewidth]{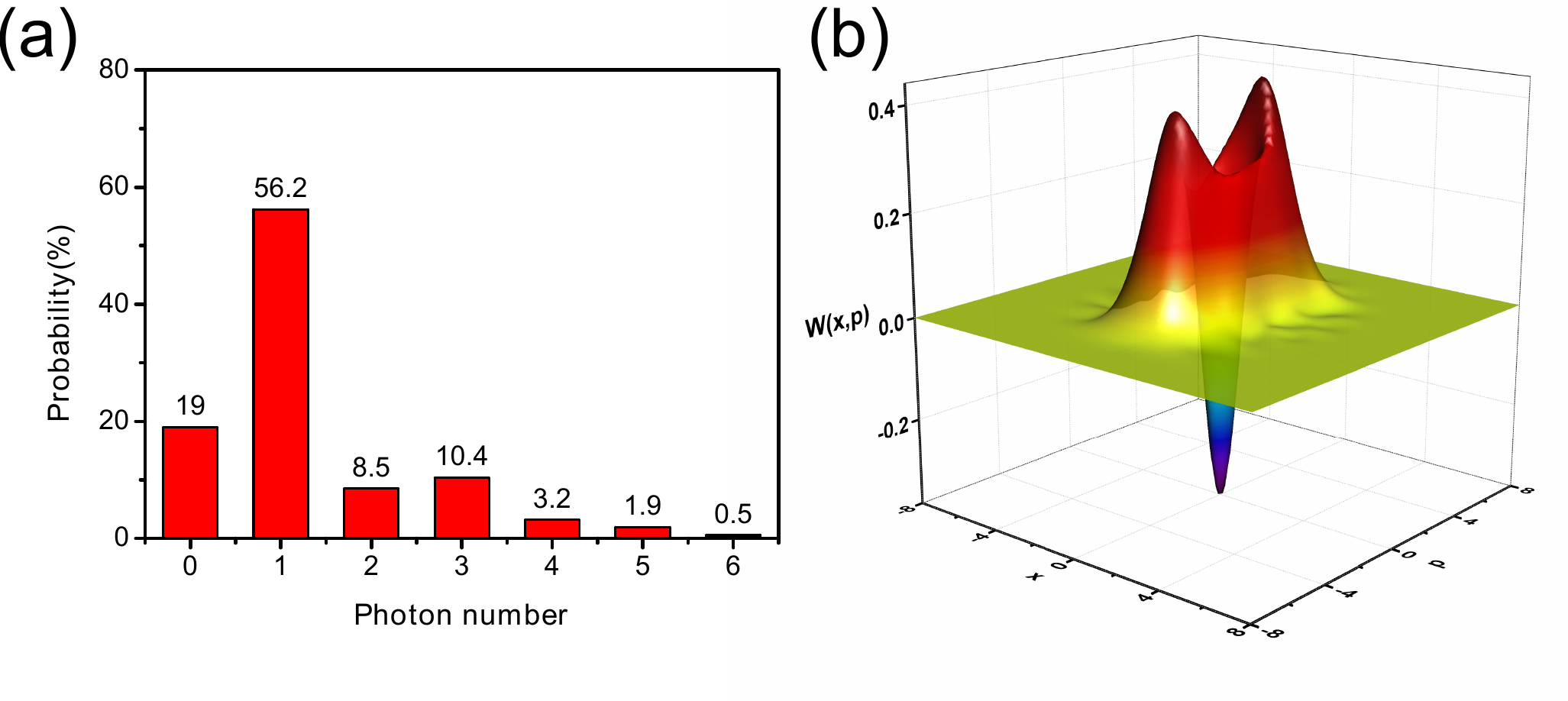}}
\caption{\textbf{Coherent state superposition (Schr\"{o}dinger kitten state).} (a) Diagonal elements of the reconstructed density matrix without correction from detection losses. (b) Corresponding Wigner function. x and p denote quadrature components.}
\label{fig4}
\end{figure}

The threshold of the type-I OPO is about 50 mW. In order to observe strong squeezing, we perform measurements close to threshold, i.e. with a pump power of 40 mW, and at an analysis frequency of 5 MHz. As shown in Figure \ref{fig3}a, the measured squeezing is -10.5 $\pm$ 0.5 dB relative to the shot noise (without any corrections, 16 $\pm$ 1 dB if corrected for detection loss and electronics noise), and the anti-squeezing is 19 $\pm$ 0.5 dB. The full noise spectra from 0 to 50 MHz at pump power of 40 mW and 5 mW is shown in Figure \ref{fig3}b. At a pump power of 5 mW, the values of squeezing and anti-squeezing are nearly the same, leading to a state with a purity close to unity. This high-purity squeezed vacuum state is used to prepare the CSS state. The tomographic reconstruction of the heralded state is given in Figure \ref{fig4}, where the diagonal elements of the reconstructed density matrix and the corresponding Wigner function are displayed.

\section{Discussion}
The conditional preparation technique presented here is always an interplay between the initial bipartite resource and the measurement performed by the heralding detector. These two components strongly influence the quantum properties of the generated state.

First, the purity of the prepared states strongly depends on the one of the initial resource, thus a good OPO is required. What is a \textit{good} OPO? It is a device for which the escape efficiency $\eta$ is close to unity. The parameter $\eta$ is given by the ratio of the transmission of the output coupler $T$, and the sum of this transmission and the intra-cavity losses (coming from scatterings or absorption in the crystal), $L+T$. For a given $L$, the transmission of the output should be increased, at the expense of a threshold increasing quadratically with this transmission. The escape efficiency directly defines the maximal amount of squeezing that can be obtained close to threshold. Here, the escape efficiency is about 96\% for both OPO. For the conditional preparation, the OPO is then operated far from the threshold to guarantee a high purity.

Another factor comes from the heralding single-photon detection. First of all, current single-photon detectors are mostly on/off detectors, only able to herald the detection of at least 1 photon. For this reason, it is crucially important to be in a regime where the probability to have two photons in the conditioning path is very low compare to the probability to have one photon. Secondly, detectors can be noisy. Such events do not herald the generation of the targeted state and result in a mixture of the heralded state and the initial resource. Specifically, they would lead to an admixture of vacuum in the single-photon preparation or of squeezed vacuum in the CSS preparation. In our experiment, we use a superconducting single-photon detector to limit this contribution. The dark noise is around a few hertz (whereas single-photon count rate is tens of kHz).

The method presented here enables the reliable generation of non-Gaussian states with a high-fidelity, mainly limited by the losses in the detection due to the close to unity escape efficiency of the OPO. Furthermore, the well-controlled spatiotemporal mode in which they are generated will facilitate their use in subsequent protocols where such states may need to interfere with other optical resources, e.g. in optical gate implementations \cite{28} or complex state engineering \cite{29}.\\

\section*{Acknowledgments}
This work is supported by the ERA-NET CHIST-ERA (QScale) and by the ERC starting grant HybridNet. F. Barbosa acknowledges the support from CNR and FAPESP, and K. Huang the support from the Foundation for the Author of National Excellent Doctoral Dissertation of China (PY2012004) and the China Scholarship Council. C. Fabre and J. Laurat are members of the Institut Universitaire de France.

\end{document}